\documentstyle[preprint,aps]{revtex}

\begin{document}
\draft
\title{Quantum Mechanics as a Classical Field Theory}
\author{A. C. de la Torre and A. Daleo
}
\address{Departamento de F\'{\i}sica,
 Universidad Nacional de Mar del Plata\\
 Funes 3350, 7600 Mar del Plata, Argentina\\
dltorre@mdp.edu.ar
}
\date{\today}
\maketitle
\begin{abstract}
The formalism of quantum mechanics is presented in a way that
its interpretation as a classical field theory is emphasized.
Two coupled real fields are defined with  given equations of motion.
Densities and currents associated to the fields are found with their 
corresponding conserved quantities. The behavior of these quantities 
under a galilean transformation suggest the association of the 
fields with
a quantum mechanical free particle. An external potential is introduced
in the Lagrange formalism. The description is equivalent to the 
conventional Schr\"odinger equation treatment of a particle. 
We discuss the attempts to build
an interpretation of quantum mechanics based on this
scheme. The fields become the primary onthology of the theory and the 
particles appear as emergent properties of the fields. These interpretations 
face serious problems for systems with many degrees of freedom.
\end{abstract}
\narrowtext
\noindent
I. INTRODUCTION
\par
After almost one century that Planck and Einstein made the first quantum
postulates\cite{plk,ein} and after 70 years that the mathematical formalism
of quantum mechanics was established\cite{vneum},
the  challenge posed by quantum mechanics is still open.
Until now, no completely satisfactory interpretation of quantum mechanics 
has been found and we can still say today that ``nobody  understands 
quantum mechanics''\cite{fey}. The lack of interpretation was compensated
by the development of an extremely precise and esthetic mathematical 
formalism; we do not know {\em what} quantum mechanics is but
we know very well {\em how} it works. The development of the very 
successful axiomatic formalism had the consequence that many physicists
where satisfied with the working of quantum mechanics and did no longer 
tried to understand it. This attitude was favored by the establishment 
of an orthodox instrumentalist ``interpretation'' which, if we are 
allowed to put it in a somewhat oversimplified manner, amounts to say 
``thou shall  not try to understand quantum mechanics''.  Only a few
authorities like Einstein, Schr\"odinger, Plank, could dare not to accept
the dogma and insist in trying to understand quantum mechanics\cite{jam}.
Fortunately the situation has changed and today it is an acceptable 
subject to search for an interpretation of quantum mechanics. The roots 
for this change are found in the pioneering work of Einstein Podolsky and 
Rosen\cite{epr} which pointed out to some peculiar correlations in the 
theory; followed by the work of Bell\cite{bel} that established 
conditions for the existence of those correlations in nature and 
finally, the experimental evidence for their existence\cite{exp}.
Although no definite interpretation of quantum mechanics has been 
found, we have made some progress in the understanding of the 
quantum world. There exist correlations among the observables of a 
quantum system that can not be explained by some classical effects\cite{dlt}. 
These correlations appear always among  noncommuting 
observables and also appear in some cases among {\em commuting} ones  
as is the case, for instance, in nonlocal or noseparable states of
quantum systems. It is no longer possible to think that the 
observables assume values independent of the context, that is, of the 
values assigned to other (commuting) observables\cite{per}.
\par
It is impossible to assign the established features of the 
quantum systems to a classical particle. It is impossible to develop 
an image for a particle having those properties. Therefore if we 
want to  understand quantum mechanics and if we 
keep in mind the image of a particle, we are in troubles. 
In this work we will see that it is possible to develop an image, 
not of a particle but of a field, that is compatible with the properties 
of a quantum system.
Guided by  our sense perception we may have the tendency to assign somehow 
a higher priority to the particles than to  the fields. 
Indeed we have a clear  sense perception  for macroscopic particles
but we have no ``feeling'' for the fields (only after a long intellectual 
development did we realize that ``light'' is a field). 
So usually we think of
particles as having onthological entity, that is, as really existent, 
whereas fields are some mathematical construction {\em associated} 
to particles like, for instance, the electric or gravitational fields 
associated to charged or massive particles.
Classical electromagnetism give us however some indication that this 
hierarchy may be incorrect. True, electric fields are a property 
of {\em charged} particles but we find, through Maxwell equations, 
that time varying electric fields can exist without  charged 
particles associated to them. 
We have therefore no deep reason to think that particles are  
``more fundamental'' than fields. It could even be the other way 
round: we will see that 
quantum mechanics is 
simpler if we consider that the fundamental entities are the fields 
instead of taking the particles as primary objects. Quantum mechanics, 
if considered as a field theory, is no more weird than electrodynamics
but if we take it as a theory of particles, very strange, 
 unnatural and contradictory things must 
be introduced. For instance we must assign to a particle, a paradigm of 
a localized thing, a nonlocal quality. Movement and position become 
incompatible although movement {\em is} the change of position for all 
images that we can make of a particle. After having understood  the 
mathematical and physical properties of the fields, we can recognize 
that some of its features can be  associated to some properties of 
particles. In this way we may recover the particles as  being emergent 
properties of the fields.
One further advantage of considering nonrelativistic 
quantum mechanics as a field theory is that this paves the way 
towards {\em  relativistic} quantum field theory. 
\par
In this work we will see a scheme  that avoids the counterintuitive 
structures that 
appear when the quantum behavior of particles is presented. The main idea 
of the work is to present quantum mechanics as a field theory with 
acceptable features, no more abstract than the ones found in classical 
electromagnetism. We will see however that this  revival
of  interpretations of quantum mechanics that assign onthological 
reality to the fields must face some severe problems. 
In  section II, the mathematical structure of a 
field theory is presented and associated, in section III, to a free 
physical system.
A potential is easily introduced in section IV and contact with 
the conventional formalism of quantum mechanics is made in section V. 
In section VI the possibility to build an interpretation of quantum 
mechanics based on this scheme is discussed.
\par
\noindent
II. MATHEMATICAL FEATURES
\par
Let us assume a physical system represented by two coupled real fields.  
From their equations of motion we can find all 
relevant properties of the fields, and associate them to physical concepts.
Let $A(x,t)$ and $B(x,t)$ denote the fields in the simplest case of 
a one dimensional space and time. These fields have no external sources 
like the electric charges and currents for the electromagnetic fields, 
but each field acts as a source 
for the polarization of the other field. 
This coupling through the polarization becomes clear in a discrete 
simulation of these fields in a 
lattice\cite{dlt1}, where particles and antiparticles associated to the 
fields are created at neighboring sites in each step of time evolution.
Let the time evolution of the fields be determined by the equations 
\begin{eqnarray}
 \partial_{t}A(x,t) =& - &\partial_{x}^{2}B(x,t) \ ,
\nonumber \\
\partial_{t}B(x,t) =&  &\partial_{x}^{2}A(x,t) \ .
\end{eqnarray}
We will use  $\partial_{x}^{n}A(x,t)$ to denote the $n$-th order partial 
derivative with respect to $x$ and similarly for the time derivatives. 
In these equations we have suppressed a constant in order to take the 
fields $A$ and $B$, as well as the space-time variables, as dimensionless. 
For the moment we are only interested in the mathematical structure of the 
fields. All constants and dimensions needed to make contact with physical 
reality can be introduced later.
These two equations play the similar r\^ole as Maxwell equations for 
the electric and magnetic fields, but are however much simpler. 
 By direct substitution we can prove the following result:
\par
If $A(x,t)$ and $B(x,t)$ are solutions of the Eqs. 1 then
$\partial_{x}^{n}A(x,t)$ and $\partial_{x}^{n}B(x,t)$ $\forall n$ are
also solutions. The same can be said for the time derivatives 
$\partial_{t}^{m}A(x,t)$ and $\partial_{t}^{m}B(x,t)$ $\forall m$. 
\par
Since the Eqs. 1 are linear, all linear combinations of solutions are 
also solutions. Therefore if $A(x,t)$ and $B(x,t)$ are solutions of 
the Eqs. 1, then
$\sum_{n} \lambda_{n}\partial_{x}^{n}A(x,t)$ and 
$\sum_{n} \lambda_{n}\partial_{x}^{n}B(x,t)$ are also solutions. As a 
special case we may choose $\lambda_{n}= L^{n}/n!$ and the 
summations become the Taylor expansion of $A(x+L,t)$ and $B(x+L,t)$.
Therefore the solutions of Eqs. 1 are invariant under space translations
$x\rightarrow x+L$. In the same way we can prove that the solutions are 
also invariant under a time translation $t\rightarrow t+T$. These two 
results also follow directly from inspection of Eqs. 1.
Another symmetry of the solutions that can be proven by direct 
substitution is that, if $A(x,t)$ and $B(x,t)$ are solutions of 
Eqs. 1 then
\begin{eqnarray}
 A'(x,t) =& c A(x,t) - s B(x,t) \ ,
\nonumber \\
 B'(x,t) =& s A(x,t) + c B(x,t) \ ,
\end{eqnarray}
where $c$ and $s$ are arbitrary real constants, are also solutions.
An interesting question is whether the constants $c$ and $s$ can become
functions  $c(x,t)$ and $s(x,t)$. Indeed, it can be easily proven that 
$A'(x,t)$ and $B'(x,t)$ given below are also solutions.
\begin{eqnarray}
 A'(x,t) =& \cos\left( \frac{v}{2}(x-\frac{v}{2}t)\right) A(x-vt,t) - 
\sin\left( \frac{v}{2}(x-\frac{v}{2}t)\right) B(x-vt,t) \ ,
\nonumber \\
 B'(x,t) =& \sin\left( \frac{v}{2}(x-\frac{v}{2}t)\right) A(x-vt,t) + 
\cos\left( \frac{v}{2}(x-\frac{v}{2}t)\right) B(x-vt,t) \ ,
\end{eqnarray}
where $v$ is an arbitrary constant. This solution is interesting 
because it shows how the fields behave in a galilean 
transformation $x\rightarrow x'=x-vt$, $t\rightarrow t'=t$, if 
we require that the  equations 
of motion in Eqs. 1 remain invariant. A similar situation is found 
in electrodynamics where the requirement of invariance of Maxwell 
equations under a Lorentz transformation mixes the electric and 
magnetic fields. In this nonrelativistic case we ask for invariance 
under a galilean transformation since Eqs. 1 are clearly not invariant 
under Lorentz transformation because of the different treatment of time and 
space variables. The equations above, considered as an active 
transformation, show how to boost the fields at an initial time ($t=0$)
with a velocity $v$.
\par
We will see now that Eqs. 1 fully determine the time evolution of the fields. 
That is, given the fields at some time, say $t=0$, we can determine the 
fields for all other times. In order to see this, we first derive an 
expression for all time derivatives of the fields in terms of their 
space derivatives. The result for the $A$ field is:
\begin{equation}
\partial_{t}^{n}A = \left\{ 
\begin{array}{ll}
(-1)^\frac{n}{2} \partial_{x}^{2n}A        &n=0,2,4,\cdots \\
-(-1)^\frac{n-1}{2} \partial_{x}^{2n}B  \  &n=1,3,5,\cdots
\end{array}
\right.\ .
\end{equation}
The result for $B$ is similar, except for a sign change when $n$ is 
odd, and can be obtained from the above equation by the replacement 
$A\rightarrow B$ and $B\rightarrow -A$. The proof of these equations 
follows by iterative time derivatives of Eqs. 1. Now we can use these
time derivatives in a Taylor expansion around the value $t=0$. 
That is, for the $A$ field we have,
\begin{equation}
A(x,t) = \sum_{n=0}^{\infty} 
\frac{t^{n}}{n!}
\partial_{t}^{n}\!A |_{t=0}
=\sum_{n=0,2,4,\cdots}^{\infty} 
\frac{t^{n}}{n!} (-1)^\frac{n}{2}\partial_{x}^{2n}\!A(x,0) -
\sum_{n=1,3,5,\cdots}^{\infty}
\frac{t^{n}}{n!}(-1)^\frac{n-1}{2} \partial_{x}^{2n}\!B(x,0) \ .
\end{equation}
We can nicely express this result if we notice that the sums above 
correspond to the series expansion of sine and cosine. Therefore we 
obtain the {\em formal} expression, also for the $B$ field,
\begin{eqnarray}
 A(x,t) =& &\cos\left( t\partial_{x}^{2}\right) A(x,0) - 
\sin\left( t\partial_{x}^{2}\right) B(x,0) \ ,
\nonumber \\
 B(x,t) =& &\sin\left( t\partial_{x}^{2}\right) A(x,0) + 
\cos\left( t\partial_{x}^{2}\right) B(x,0) \ .
\end{eqnarray}
\par
Now we want to find some conserved properties of the fields. These 
conserved properties are candidates for a physical interpretation. 
In the search for conserved quantities, we must eliminate the space 
dependence. To do this we can integrate the fields and functions of 
them in all space. In order to have such integrals  
mathematically well defined, we  impose the supplementary condition 
that the fields should be confined. This means that the fields and 
all their derivatives must vanish at infinity.
\begin{equation}
 \partial_{x}^{n}A(\pm \infty,t) =
\partial_{x}^{n}B(\pm \infty,t)=0\ ,\ \forall n=0,1,2,\cdots\ .
\end{equation}
Actually we need the stronger condition that they should vanish
faster than any power of $x$.
\par
Before we identify the conserved quantities, we need a couple of definitions.
Given two fields $A(x,t)$ and $B(x,t)$, we define their associated
$n^{th}$ {\em order  density} ${\cal M}_{n}(x,t)$ and 
$n^{th}$ {\em order  current} ${\cal P}_{n}(x,t)$, (shortly n-density 
and n-current) by
\begin{eqnarray}
 {\cal M}_{n}(x,t) &=&  (\partial_{x}^{n}\!A)^{2} +(\partial_{x}^{n}\!B)^{2}  \ ,
\nonumber \\
 {\cal P}_{n}(x,t) &=&  \partial_{x}^{n}\!A\,\partial_{x}^{n+1}\!\!B
-\partial_{x}^{n}\!B\,\partial_{x}^{n+1}\!\!A\ ,\ n=0,1,2\cdots\ .
\end{eqnarray}
The reason for the names chosen is that if $A(x,t)$ and $B(x,t)$ 
are solutions of Eqs. 1 then these quantities obey a 
{\em continuity} equation,
\begin{equation}
\partial_{t}{\cal M}_{n}(x,t) +
2 \partial_{x}{\cal P}_{n}(x,t) =0 \ .
\end{equation}
This can be proven directly performing the derivatives and replacing
from Eqs. 1. 
We can now prove that the space integrated n-densities and n-currents
are conserved. That is, the quantities $M_{n}$ and $P_{n}$ defined as
\begin{eqnarray}
M_{n} &= 
 \int^{\infty}_{-\infty}\, {\cal M}_{n}(x,t)\, dx \nonumber \\
P_{n} &= 
 \int^{\infty}_{-\infty}\, {\cal P}_{n}(x,t)\, dx   \ ,
\end{eqnarray}
are such that
\begin{equation}
\partial_{t} M_{n} = 0 \ ,\ \partial_{t} P_{n} = 0 \ .
\end{equation}
These equations can be proven performing the time derivative of the products
in Eqs. 8 and replacing them with Eqs. 1 and finally doing appropriate 
integrations by parts with vanishing border terms due to Eq. 7. However,
the first of these equations follows easier from the continuity equation:
\begin{equation}
\partial_{t} M_{n} = 
\int^{\infty}_{-\infty}\, \partial_{t}{\cal M}_{n}(x,t)\, dx 
=-2\int^{\infty}_{-\infty}\, \partial_{x}{\cal P}_{n}(x,t)\, dx  
\left.
=-2 \, {\cal P}_{n}(x,t) \right|^{\infty}_{-\infty}
= 0 \ ,
\end{equation}
where, again, the last term vanishes because of Eq. 7.
For later reference, it is convenient to perform $n$ integrations by parts
of the densities and currents in Eqs. 10. The border terms in all these 
integrations vanish and we get the results
\begin{eqnarray}
M_{n} &= (-1)^{n}
 \int^{\infty}_{-\infty}\, 
\left( A\partial_{x}^{2n}\!A +B\partial_{x}^{2n}\!B\right)
\, dx \nonumber \\
P_{n} &= (-1)^{n}
 \int^{\infty}_{-\infty}\,
\left( A\,\partial_{x}^{2n+1}\!\!B
-B\,\partial_{x}^{2n+1}\!\!A\right) \, dx   \ .
\end{eqnarray}
\par

\noindent
III. PHYSICAL FEATURES
\par
In the last section we have identified several mathematical features
of the fields $A(x,t)$ and $B(x,t)$ determined by Eqs. 1. Now we want 
to investigate what physical systems these fields may describe. A 
natural thing to do is to associate the conserved properties of the 
fields with some permanent features of the physical system. 
\par
We have seen that if we know the fields at a particular time, then 
the solutions are determined for all times. However there is a 
large class of possible  initial conditions, producing different 
solutions to the field equations. We will identify these different 
solutions with {\em different states} of the same physical system. 
All these solutions share the property of invariance under space 
and time translation $x\rightarrow x+L$, $t\rightarrow t+T$.
That is, it is irrelevant {\em where and when} the system is considered.
Such a system is {\em free}, that is, noninteracting with any external
system.
\par
The fields have an inescapable and inherent time 
dependence. That is, there are no {\em static} $A(x,t)$ and $B(x,t)$ 
fields. Clearly, if the fields have no time dependence, that is if
$\partial_{t}\!A=\partial_{t}\!B=0$,  the only solution consistent with
the asymptotic conditions of Eq. 7 is $A(x,t)=B(x,t)=0$. 
In electromagnetism we have static fields that are generated by
static electric charges and currents. Our field equations have no 
such sources and therefore we do not have static solutions.
The fields $A$ and $B$ must have then a time dependence. There are 
however solutions, with time varying fields, such that the 
n-densities and n-currents are time independent. Let us call them
{\em stationary} solutions. From  Eqs. 8 we can find that 
the condition $\partial_{t}{\cal M}_{n}(x,t)=0$ is satisfied if
\begin{eqnarray}
 \partial_{t}A(x,t) =& E B(x,t) \ ,
\nonumber \\
\partial_{t}B(x,t) =& -E A(x,t) \ ,
\end{eqnarray}
where $E$ is some constant; and $\partial_{t}{\cal P}_{n}(x,t)=0$ is 
valid if
\begin{eqnarray}
 \partial_{x}^{2}A(x,t) &= -E A(x,t) \ ,
\nonumber \\
\partial_{x}^{2}B(x,t) &= -E B(x,t) \ .
\end{eqnarray}
To reach this last condition we have used Eqs. 1. These two sets of 
equations are the sufficient (necessary?) conditions for the stationary 
states and are consistent with Eqs. 1. Indeed, the second set can be 
obtained using the first one in Eqs. 1. Each stationary solution is then 
characterized by a real constant $E$ (we will see later that $E>0$).
These stationary solutions are linearly independent because
linear combinations of stationary solutions are {\em not} 
stationary solutions. Indeed, if $E_{1}$ and $E_{2}$ characterize 
the solutions $(A_{1},B_{1})$ and $(A_{2},B_{2})$, one can prove that 
$(c_{1}A_{1}+c_{2}A_{2},c_{1}B_{1}+c_{2}B_{2})$ is not a stationary solution
(although it is of course a solution to Eqs. 1).
In order to study the properties of these 
solutions it is very useful to notice, from Eqs. 15, that 
the differential operator $\partial_{x}^{2}$, {\em when applied to the 
stationary solutions}, can be simply replaced  by the constant $-E$. With
this we can immediately give the stationary state in terms of 
some initial  stationary state, replacing 
$\partial_{x}^{2}\rightarrow -E$ in Eqs. 6:
\begin{eqnarray}
 A(x,t) =& &\cos\left(E t\right) A(x,0) +
\sin\left( E t\right) B(x,0) \ ,
\nonumber \\
 B(x,t) =&-&\sin\left( E t\right) A(x,0) + 
\cos\left(E t\right) B(x,0) \ .
\end{eqnarray}
In the ``$A,B$ plane'', the stationary states, at fixed $x$, rotate in 
a circle of radius
${\cal M}_{0}$ with constant angular velocity $E$. With respect to time, 
the stationary solutions oscillate with (time) frequency $Et$ and with 
respect 
to space they oscillate with (space) frequency $\sqrt E x$, as can be 
seen from the  solutions of Eqs. 15. Here we must impose the condition
$E>0$ in order to avoid divergent solutions.
\par
In the stationary states, the n-densities and n-currents have no time 
dependence. What about their spatial dependence? For the n-currents the 
answer is immediately given by the continuity equation 9
\begin{equation}
\partial_{t}{\cal M}_{n}(x,t)=0\  \longrightarrow
\ \partial_{x}{\cal P}_{n}(x,t)=0 \ .
\end{equation}
The n-currents in the stationary states are therefore constants independent 
of time and space. To find the value of these constants we can use 
the replacement $\partial_{x}^{2}\rightarrow -E$ in Eqs. 8 to find
\begin{equation}
{\cal P}_{n}=E{\cal P}_{n-1}\ \longrightarrow \ 
{\cal P}_{n}=E^{n}{\cal P}_{0}\ .
\end{equation}
The n-densities are, as seen, time independent but in general they 
may depend on $x$.
With the same replacement $\partial_{x}^{2}\rightarrow -E$
we can prove that (dropping explicitly the time variable $t$)
\begin{equation}
{\cal M}_{n}(x)= \left\{ 
\begin{array}{ll}
 E^{n}{\cal M}_{0}(x)  \   &n=0,2,4,\cdots \\
E^{n-1}{\cal M}_{1}(x)  \  &n=1,3,5,\cdots
\end{array} 
\right. \ .
\end{equation}
Furthermore, ${\cal M}_{1}$ can be given in terms of ${\cal M}_{0}$ because
$\partial_{x}{\cal M}_{1} 
= \partial_{x}((\partial_{x}A)^{2} + (\partial_{x}B)^{2}) 
= 2 \partial_{x}A\partial_{x}^{2}A + 2\partial_{x}B\partial_{x}^{2}B
=-E(2 A\partial_{x}A + 2B\partial_{x}B)
=-E\partial_{x}((A)^{2}+(B)^{2})
=-E\partial_{x}{\cal M}_{0}\ $; therefore ${\cal M}_{1}=E(C-{\cal M}_{0})$ 
with an arbitrary constant $C$. This constant must be such that
$C>\max{\cal M}_{0}$ to guarantee positivity of ${\cal M}_{1}$.
\par
These stationary solutions oscillate for all time and in all space. They are
similar to the electromagnetic plane wave solutions to Maxwell's equations. 
The same  happens here as with the electromagnetic plane waves: these 
solutions can only be considered to represent approximately a physical 
system.  Strictly speaking these solutions are unphysical because they 
imply an infinitely extended system and the conserved quantities of 
Eqs. 10 are meaningless. However they are useful to 
represent  approximate situations and, most important, these solutions
are linearly independent and therefore
can be used in superposition in order to construct 
{\em wave packets} with any desirable shape and extension.
There is however an 
important difference between the wave packets solutions to Maxwell equations 
and the propagating ``wave packets'' solutions to our Eqs. 1. There exist 
electromagnetic pulses or packets of arbitrary shape that propagate  
without changing the shape. 
This is not possible in our case. Indeed, we can show that our Eqs. 1 
do not admit solutions of the type $A=f(x+vt),\ B=g(x+vt)$ where $f$
and $g$ are {\em arbitrary} (differentiable) functions. Only sine and 
cosines are acceptable, and these are the stationary solutions 
already mentioned. Therefore the shape of our field distributions 
must change during the time evolution. In other words, 
the fields representing a quantum system described by Eqs. 1 
are {\em dispersive}: localization and shape of the distributions are 
not a permanent feature of quantum systems. 
\par
We will see now, for arbitrary states, how the conserved quantities 
behave under the 
transformations of the fields. Let us first consider the 
transformation  described by  Eqs. 2. If the fields
$A(x,t)$ and $B(x,t)$   have the n-densities and n-currents
${\cal M}_{n}(x,t)$ and  ${\cal P}_{n}(x,t)$
and their corresponding conserved quantities
$M_{n}$ and  $P_{n}$, then the transformed fields
$A'(x,t)$ and $B'(x,t)$ given by Eqs. 2 will correspond to
${\cal M}'_{n}(x,t)=(c^{2}+s^{2}){\cal M}_{n}(x,t)$ and  
${\cal P}'_{n}(x,t)=(c^{2}+s^{2}){\cal P}_{n}(x,t)$ and also,
$M'_{n}=(c^{2}+s^{2})M_{n}$ and  $P'_{n}=(c^{2}+s^{2})P_{n}$.
Therefore this transformation produces just a change of scale
that can be canceled by a numerical factor multiplying the fields. 
Furthermore, if the constants are such that $c^{2}+s^{2}=1$ then 
the transformation is irrelevant for the physical quantities 
represented by the n-densities and n-currents and their conserved quantities.
\par
The next transformation of Eqs. 3 gives much more information about the 
physical nature of the conserved quantities. This transformation 
represents either the observation of the same physical system from 
a reference frame moving with velocity $-v$ or alternatively, the 
same physical system ``boosted'' with a velocity $v$. Let us consider 
first the 0-density. Although the fields are mixed in this 
transformation, this density remains unchanged in shape and is just 
boosted with velocity $v$.
\begin{equation}
{\cal M}'_{0}(x,t)=A'^{2}(x,t)+B'^{2}(x,t)={\cal M}_{0}(x-vt,t)\ .
\end{equation}
This suggests that the 0-density ${\cal M}_{0}(x,t)$ represents the
{\em space localization} of the physical system described by the fields
$A(x,t)$ and $B(x,t)$. We can always scale the field such that the 
conserved quantity  associated to the 0-density takes the value 
$M_{0}=1$. With this normalization, the 0-density can be thought as a 
distribution of localization.
Let us consider now the 0-current ${\cal P}_{0}(x,t)$. Using the 
transformation of Eqs. 3 we obtain, after straightforward manipulations,
\begin{equation}
{\cal P}'_{0}(x,t)
={\cal P}_{0}(x-vt,t) +\frac{v}{2} {\cal M}_{0}(x-vt,t)\ ,
\end{equation}
and the corresponding conserved quantities, assuming the normalization
$M_{0}=1$,
\begin{equation}
P'_{0}= P_{0} +\frac{v}{2} \ .
\end{equation}
 For the 1-density we get:
\begin{equation}
{\cal M}'_{1}(x,t)
={\cal M}_{1}(x-vt,t)+v{\cal P}_{0}(x-vt,t) 
+\left(\frac{v}{2}\right)^{2} {\cal M}_{0}(x-vt,t)\ ,
\end{equation}
and the corresponding conserved quantities are
\begin{equation}
 M'_{1}= M_{1}+v P_{0}+\left(\frac{v}{2}\right)^{2}\ .
\end{equation}
The conserved quantities $P_{0}$ and $M_{1}$ can be consistently 
associated to the momentum and kinetic energy of the system 
(with mass $m=1/2$)
because under a boost of velocity $v$ they are changed accordingly.
Indeed if $P_{0}=\frac{1}{2} u$ and $M_{1}=\frac{1}{4} u^{2}$ then
$P'_{0}=\frac{1}{2} (u+v)$ and $M'_{1}=\frac{1}{4} (u+v)^{2}$. For 
the remaining n-densities and n-currents we can also find their 
transformation properties in a boost but there are no remaining 
particle observables to which they can be related. Furthermore,
one must be cautious in the identification of the conserved properties 
of the quantum system with the conserved properties of a particle. 
For instance, the relation between kinetic energy and momentum valid
for classical particles is not always true for the quantum system. 
One can prove that  in general it is $M_{1} \geq P_{0}^{2} $
and not $M_{1} =P_{0}^{2} $ as one would expect for a particle of mass
$m=1/2$. Of course,  the identification of a delocalized quantum 
system, for instance in a stationary state, with a localized 
classical particle becomes very suspect.
\par
We can obtain further confirmation that the system is moving with 
momentum $P_{0}$ by calculating the time derivative of the center of the 
localization distribution defined as
\begin{equation}
X=\int^{\infty}_{-\infty} x {\cal M}_{0}(x,t)\,dx \ .
\end{equation}
The center of the distribution  moves with constant velocity
$2P_{0}$:
\begin{eqnarray}
\partial_{t}X &=\int^{\infty}_{-\infty} x \partial_{t}{\cal M}_{0}(x,t)\,dx
=-2\int^{\infty}_{-\infty} x \partial_{x}{\cal P}_{0}\,dx
\nonumber\\
&=-2x{\cal P}_{0}|^{\infty}_{-\infty}+
2\int^{\infty}_{-\infty} {\cal P}_{0}\,dx
=2P_{0} \ ,
\end{eqnarray}
where we have used the continuity equation and that the border term of 
the integration by parts  vanish because  the 
vanishing in Eq. 7 is faster than any power of $x$. 
We can use the relation between the velocity of the center of the 
distribution and the conserved momentum in order to {\em define} 
the mass of the free quantum system: $m=P_{0}/\partial_{t}X$.
Notice that with 
a similar calculation we reach the conclusion that the center of the 
distributions, corresponding to the n-densities ${\cal M}_{n}(x,t)$, all 
move with constant ``velocities'' equal to $2P_{n}$. The  fact that these 
velocities are not all equal is indicative of the dispersive nature  
in the evolution of the quantum system.
\par
We summarize:
the fields $A(x,t)$ and $B(x,t)$ determined by Eqs. 1 describe a free quantum 
system localized according to the density distribution ${\cal M}_{0}(x,t)=
A^{2}(x,t)+B^{2}(x,t)$ moving with a constant drift velocity
$2P_{0}=2\int  (A\,\partial_{x}\!B -B\,\partial_{x}\!A\ )dx $
and constant kinetic energy 
$M_{1}=\int  (\partial_{x}\!A)^{2} +(\partial_{x}\!B)^{2} dx $.
Referring to a similar classical system, 
we can denote this system as a {\em free quantum particle}. 
The shape of the distributions change in  time,  indicating that 
localization is not a permanent property of the free quantum system.
When the 
distribution ${\cal M}_{0}(x,t)$ is localized in a small region, 
the system is in a {\em particle-like state}. 
If the distribution ${\cal M}_{0}(x,t)$ is 
delocalized in a large 
region, the system is in a {\em wave-like state}. 
\par
\noindent
IV. EXTERNAL INTERACTION
\par
The field equations given in Eqs. 1 can be derived using the Euler-Lagrange 
equations applied to the following lagrangian density, a functional of the 
fields  and of their space and time derivatives,
\begin{equation}
{\cal L}=
B\,\partial_{t}\!A - A\,\partial_{t}\!B -
(\partial_{x}\!A)^{2} -(\partial_{x}\!B)^{2}  \ .
\end{equation}
As usual,  terms like $\partial_{t}\!B\,\partial_{x}\!A-   
\partial_{t}\!A\,\partial_{x}\!B$ or 
$B\,\partial_{x}\!A + A\,\partial_{x}\!B$ and other total differentials
can be added to this lagrangian density but they are irrelevant because they
make a vanishing contribution to the field equations.
From this lagrangian density we can find the canonical field momenta 
associated to the fields $A(x,t),\ B(x,t)$. It turns out that 
$A(x,t)$ and $B(x,t)$  are reciprocally the canonical momenta of each 
other (within a minus sign). With these momenta and with the lagrangian 
density we obtain the canonical hamiltonian density 
\begin{equation}
{\cal H}=
(\partial_{x}\!A)^{2} +(\partial_{x}\!B)^{2}  \ .
\end{equation}
Here again, total differential terms can be added that have no contribution
when integrated in all space because of Eq. 7. Notice that ${\cal H}=
{\cal M}_{1}(x,t)$
and therefore the integrated hamiltonian density is conserved.  
We should not naively attempt to derive the equations of motion from the 
hamiltonian formalism using the last expression. This fails because we 
are  dealing here with a 
{\em nonstandard lagrangian} linear in the ``velocities''. For this 
lagrangian we can not solve the ``velocities'' in terms of the canonical 
momenta. We will  not treat here this  constrained hamiltonian field theory. 
We have brought here the 
hamiltonian density, that can be interpreted as an energy density, 
only for the purpose of guiding us in the introduction of an interaction 
in our system. 
\par
The free system of Eqs. 1 is translation invariant and all regions of space
are equivalent. Let us break this translation symmetry. Let us assume now 
that there are some regions of space
where the energy of the system is changed by the action of an external 
field denoted by $V(x)$. We call this a {\em potential field}. The 
presence or localization of the system in the regions where the 
potential field $V(x)$ is nonvanishing, will change the energy of 
the system by an amount given by the product of the potential with the 
density of localization of the system in such a region. This last 
density is given by ${\cal M}_{0}(x,t)$. Therefore we introduce the 
interaction of the system with an external potential $V(x)$ just 
changing the hamiltonian  density to the new expression
\par
\begin{equation}
{\cal H}=
(\partial_{x}\!A)^{2} +(\partial_{x}\!B)^{2}  
+V(x)\left(A^{2} +B^{2}  \right)\ .
\end{equation}
With this new hamiltonian, that includes the interaction, 
we get a new lagrangian density
\begin{equation}
{\cal L}=
B\,\partial_{t}\!A - A\,\partial_{t}\!B -
(\partial_{x}\!A)^{2} -(\partial_{x}\!B)^{2}  
-V(x)\left(A^{2} +B^{2}  \right)\ ,
\end{equation}
and with the Euler-Lagrange equations we obtain the equations of 
motions for the fields describing a quantum particle in a potential,
\begin{eqnarray}
 \partial_{t}A(x,t) =& - &\partial_{x}^{2}B(x,t) +V(x) B(x,t)\ ,
\nonumber \\
\partial_{t}B(x,t) =&  &\partial_{x}^{2}A(x,t) -V(x) A(x,t)\ .
\end{eqnarray}
An analysis similar to the one performed for the free system could be 
now done. These equations are invariant under the transformation Eq. 2 
and an observer in a moving frame will see the potential moving, 
$V(x-vt)$, as expected. The time evolution can be determined in a 
similar way as before.
A similar study of stationary states can be done  with the replacement
$\partial_{x}^{2}-V(x)\rightarrow -E$
\par
\noindent
V. CONVENTIONAL FORMALISM
\par
Once that we have seen the advantages of the description of quantum 
mechanics as a classical field theory we must make contact with the 
conventional formalism. For this we just have 
to define a {\em complex field} $\Psi (x,t) = A(x,t) +i B(x,t)$. From 
the equations of motion for the fields $A$ and $B$, given in Eqs. 31, we 
obtain the corresponding equation for $\Psi$. Furthermore we introduce
constants in order  to recover physical dimensions for space, time, 
mass and energy, and we obtain thereby Schr\"odinger's equation,
in all its glamour:
\begin{equation}
i\hbar\frac{\partial \Psi(x,t)}{\partial t}
= - \frac{\hbar^{2}}{2 m} \frac{\partial^{2}\Psi(x,t)}{\partial x^2} 
+ V(x)\Psi(x,t)      \ .
\end{equation}
All the formal expressions found before have a corresponding 
representation in terms of the complex field. For instance
${\cal M}_{0}(x,t) = |\Psi(x,t)|^{2}$. The conserved quantities 
(when $V=0$) of Eqs. 13 correspond to the expectation value  of 
even and odd powers of an operator
\begin{eqnarray}
M_{n} &= 
 \langle\Psi,(-i\partial _{x})^{2n}\Psi\rangle \nonumber \\
P_{n} &= 
 \langle\Psi,(-i\partial _{x})^{2n+1}\Psi\rangle   \ ,
\end{eqnarray}
suggesting the association of that operator with the conserved momentum.
\par
In the derivation of Schr\"odinger equation, or of the coupled equations in
Eqs. 31, from the lagrangian Eq. 30, or its equivalent in terms of 
$\Psi$ and $\Psi^{*}$, the two fields, either $A$ and $B$ or $\Psi$ and 
$\Psi^{*}$, must be varied independently. It is not possible to write 
a lagrangian in terms of {\em only one} field that leads to 
Schr\"odinger's equation. Two fields must be taken. The necessity of two 
fields is inherent to the formalism and therefore the use of a complex 
field for quantum mechanics (or the equivalent  two fields $A$ and $B$)
is different from other cases in physics, for instance in electrodynamics, 
where complex quantities are just a convenience. This necessity of two fields
is more clearly stated in the approach presented here where 
the elegance of the Lagrange formalism becomes a very convenient 
way for introducing an 
external potential. 
\par
Another advantage of the formalism in terms of the two fields 
$A$ and $B$ instead of one complex field $\Psi$ is the possibility of
an enlightening analogy with classical mechanics.
The phase space of a classical particle (in a line) is two dimensional 
and is spanned by the {\em dynamic variables} $X$ and $P$ whereas the 
``phase space'' for the quantum system is also two dimensional 
and is spanned by the {\em dynamic fields} $A$ and $B$. This parallel
can be continued  noticing that the canonical transformation 
corresponding to a rotation in phase space 
\begin{eqnarray}
 X' =& c X - s P \ ,
\nonumber \\
 P' =& s X + c P \ ,
\end{eqnarray}
with $c^{2}+s^{2}=1$, 
is equivalent to the transformation of the fields in Eqs. 2. 
Furthermore $A$ and $B$ as well as $X$ and $P$ are canonical 
conjugate of each other (within a minus sign).
\par 
\noindent
V. INTERPRETATION
\par
Quantum mechanics looses most of the strange and awkward aspects when 
it is seen as a field theory and becomes a simple subject in  comparison 
with  the unsurpassable conceptual difficulties found when we look at it 
as a theory for particles. The  advantages of presenting 
quantum mechanics as a field theory become obvious. Encouraged by this, 
we want to face now the question of whether this is just a formal
advantage or if 
this  trick  can become an {\em interpretation} of 
quantum mechanics. 
\par
To turn this into an interpretation we must decide to choose 
the fields as the {\em primary onthology}, that is, the fields are 
the really existent objects that the theory must describe, instead of 
taking the particles as the primary objects to be described 
by quantum mechanics. In a simplified manner, we may choose between 
two extreme interpretations: either the coupled fields
$A(x,t)$ and $B(x,t)$ really exist and are determined by field equations
with quite reasonable properties, no more strange than the 
electromagnetic fields, or on the contrary, the particles really exist
and the fields are a mathematical construction corresponding to 
the probability of finding, or of localization, of such particles
that must have strange schizophrenic properties of being at 
different places at the same time, and that sometimes appear as point 
like particles but some other times as extended waves. In the first option
we just have fields as the primary onthology and
``the particles'' are not  existent objects but are a name, 
learned from  the classical physics of macroscopic object, that we may 
use to denote a set of {\em emergent properties} of the fields. 
In this interpretation, what we usually call ``an electron'' is not a 
point like particle that is detected somewhere according to a 
``probability  cloud'' but is the cloud itself. 
\par
There have been several attempts of interpretations based on different 
choices for the primary onthology. Without many details we just 
mention some of them\cite{int}.  
The well known probability interpretation of the wave function proposed by 
Max Born favors  a particle onthology. In this case the field $\Psi$
does not carry energy and is not really existent in physical space. 
Opposite to it, we can find 
Schr\"odinger interpretation proposing that only the wave function 
has objective existence. In between, we can find Madelung hydrodynamical 
interpretation, no longer considered probably because it does not clearly 
states what is precisely the fluid being described by Schr\"odinger 
equation. A hybrid
 interpretation was proposed by L. de Broglie with his ``double solution'' 
suggesting a mixed particle {\em and} field onthology. Another idea 
originated by L. de Broglie is based on a particle onthology with a pilot 
wave determining its motion. This interpretation was successfully taken by 
D. Bohm in his causal quantum mechanics (Bohmian mechanics) that has 
received much attention in the lasts years. We can safely say that no one 
of the mentioned interpretations is fully satisfactory because they all 
face some difficulties, and probably also 
no one of them is totally excluded at present. It is desirable to present 
all these alternatives instead of taking, as is unfortunately 
often done, an instrumentalist position that implies an evasion from the 
problem.
\par
Since the scheme of this work clearly favors Schr\"odinger 
interpretation, we must recall the most serious problem faced in this choice.
For a single quantum system, the formalism presented can be interpreted
in the lines shown by Schr\"odinger. However for a compound system with 
many degrees of freedom we must introduce more variables upon which the 
fields depend. The fields no longer ``live'' in physical space but in a $3N$
dimensional phase space. Furthermore, if we want to give to the fields 
an onthological character then we must solve somehow the ambiguity 
in the fields apparent in Eqs. 2. Is there a way to fix the phase 
and define the fields without ambiguity? or can we tolerate the 
ambiguity and define the primary onthology as an 
{\em equivalence class} of fields? Another item that requires much 
thought is the determination of what are precisely the observable 
properties of the fields and the meaning of the measurement process. 
The fields are, apparently, not directly observables but this would not
be a serious difficulty for the interpretation, under the consideration
that  any measurement is a complex process that involves macroscopic
apparatus. The electron, as a particle in the conventional 
interpretation of quantum mechanics (whatever that is), is also, 
as is the case for the fields, not a direct observable. 
\par
\noindent
VI. CONCLUSIONS
\par
It has been shown that the consideration of quantum mechanics as a 
classical field theory can have significant  advantages. 
In this way, very confusing concepts like, for instance, 
the probability of observing, or of existence, of a particle 
somewhere, or the incompatibility  between position and momentum, 
are avoided. Quantum mechanics, as a field theory, is not 
stranger than classical electrodynamics. 
The extension of this work to the realistic three 
dimensional space is trivial and can be performed by the substitutions
$x\rightarrow \bf r$ and $\partial_{x} \rightarrow  \nabla$.
\par
The ideas presented in this work suggest a revival of the 
interpretations of quantum mechanics based on a field onthology, similar 
to the original interpretation proposed by Schr\"odinger. These 
interpretations have, however, serious difficulties
in the case of a system with many degrees of freedom 
because the dimension of  the space where the fields have support is 
not equal to the dimension of physical space. 
\par
We would like to acknowledge helpful discussions with H. M\'artin and 
P. Sisterna.
This work has received partial support from ``Consejo 
Nacional de Investigaciones Cient\'{\i}ficas y T\'ecnicas 
(CONICET), Argentina.

\end{document}